\documentclass[conference]{IEEEtran}
\usepackage[latin1]{inputenc} 
\usepackage{amsmath,amssymb}
\usepackage{anyfontsize}
\usepackage{svg}
\usepackage{booktabs}
\setsvg{svgpath=}
\usepackage{arydshln} 
\usepackage{fancyhdr}

\fancypagestyle{mahmood}{%
   \fancyhf{} 
   
   \fancyhead[C]{Published as a conference paper at the IEEE European Signal Processing Conference (EUSIPCO), 2018}
}%

\makeatletter
\let\ps@IEEEtitlepagestyle\ps@mahmood
\makeatother

\title{Fast and Accurate Multiclass Inference for MI-BCIs Using Large Multiscale Temporal and Spectral Features}
\begin{document}  

\author{\IEEEauthorblockN{Michael Hersche\IEEEauthorrefmark{1}, Tino Rellstab\IEEEauthorrefmark{1}, 
Pasquale Davide Schiavone\IEEEauthorrefmark{1},
Lukas Cavigelli\IEEEauthorrefmark{1},
Luca Benini\IEEEauthorrefmark{1}\IEEEauthorrefmark{3},
Abbas Rahimi\IEEEauthorrefmark{1}\IEEEauthorrefmark{2}
}
\IEEEauthorblockA{\IEEEauthorrefmark{1}Integrated System Laboratory, ETH Zurich, Switzerland. \IEEEauthorrefmark{3}DEI, University of Bologna, Italy.}
\IEEEauthorblockA{\IEEEauthorrefmark{2}EECS Department, 
University of California, Berkeley.}
\IEEEauthorblockA{Emails: \{herschmi, tinor\}@ethz.ch, \{pschiavo, cavigelli, lbenini, abbas\}@iis.ee.ethz.ch}
}

\maketitle

\begin{abstract}
Accurate, fast, and reliable multiclass classification of electroencephalography (EEG) signals is a challenging task towards the development of motor imagery brain--computer interface (MI-BCI) systems.
We propose enhancements to different feature extractors, along with a support vector machine (SVM) classifier, to simultaneously improve classification accuracy and execution time during training and testing.
We focus on the well-known common spatial pattern (CSP) and Riemannian covariance methods, and significantly extend these two feature extractors to multiscale temporal and spectral cases. 
The multiscale CSP features achieve 73.70$\pm$15.90\% (mean$\pm$ standard deviation across 9 subjects) classification accuracy that surpasses the state-of-the-art method~\cite{parallel_CNN}, 70.6$\pm$14.70\%, on the 4-class BCI competition IV-2a dataset.
The Riemannian covariance features outperform the CSP by achieving 74.27$\pm$15.5\% accuracy and executing 9$\times$ faster in training and 4$\times$ faster in testing.
Using more temporal windows for Riemannian features results in 75.47$\pm$12.8\% accuracy with 1.6$\times$ faster testing than CSP.
 
\end{abstract}

\begin{IEEEkeywords}
EEG, motor imagery, brain--computer interfaces, multiclass classification, multiscale features, SVM.
\end{IEEEkeywords}

\section{Introduction}
\begin{figure*}[htpb]
\subfloat[][Simplified timing scheme of one trial and the used overlapping temporal windows $T_1$--$T_{11}$.]{ 
\includesvg[width = 0.45\textwidth]{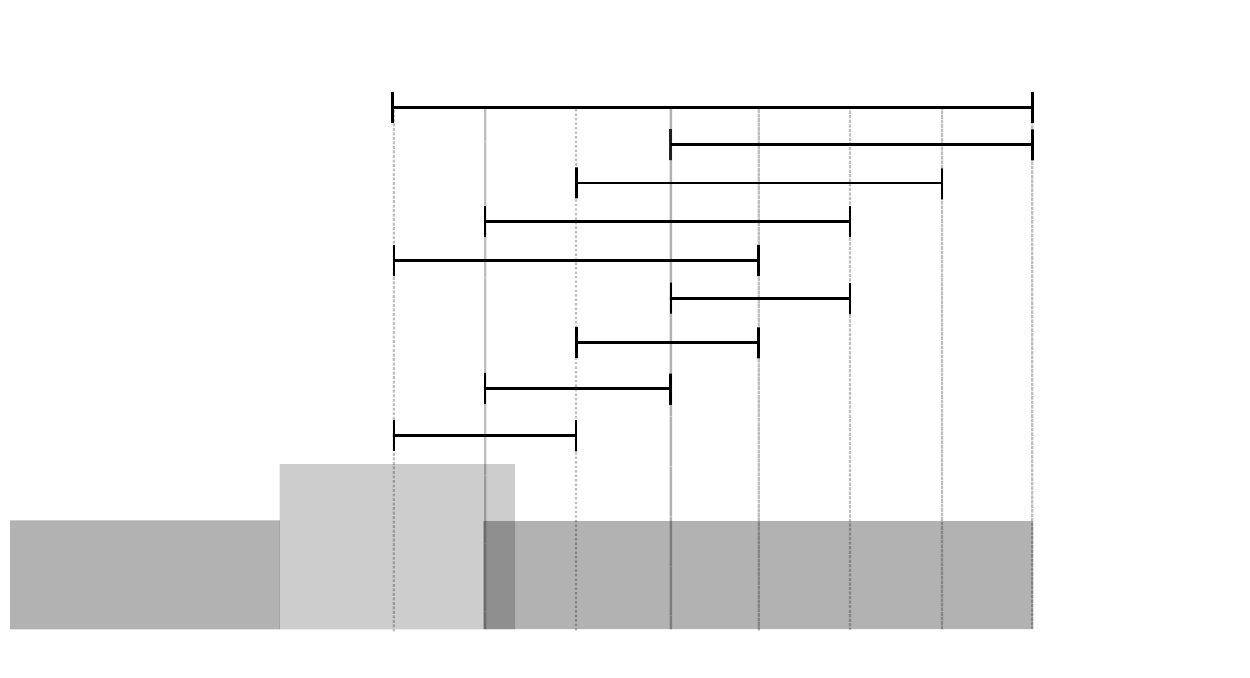}
\label{fig:temp_wind}
}
\hspace{0.05\textwidth}
\subfloat[][Overlapping spectral bands $b_1$--$b_{80}$ with bandwidths 1-32 Hz]{
\includesvg[width = 0.45\textwidth]{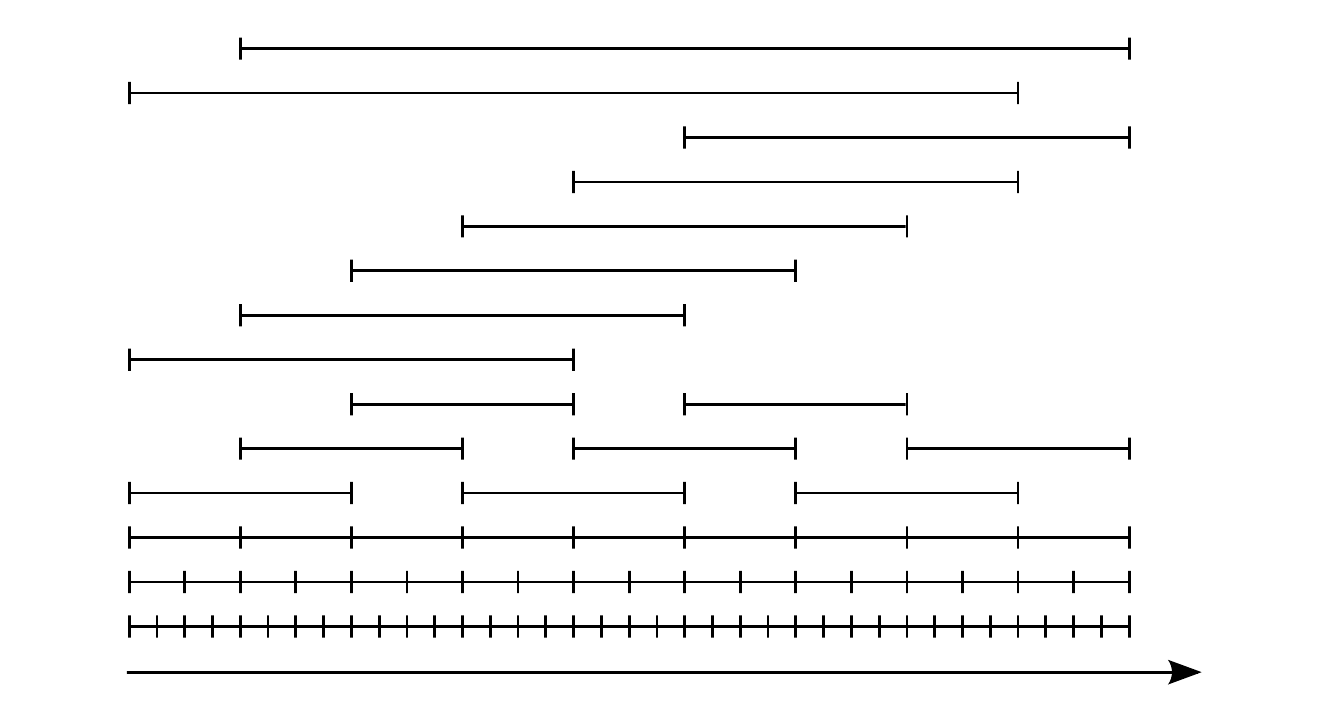}
\label{fig:freq_bands}
}
\caption{Multiscale temporal (a) and spectral (b) decomposition of the EEG signal.}
\end{figure*}
A brain--computer interface (BCI) is a system which enables communication and control between the brain and an external device.
The system aims to recognize human intentions from spatiotemporal neural activity, typically recorded non-invasively by a large set of electroencephalogram (EEG) electrodes.
There exist various applications for BCI such as motor imagery (MI), which is the cognitive process of thinking of a motion, e.g. of the left or right hand, without actually performing it. 
MI-BCI systems are designed to find patterns in the EEG signals and match the signal to the motion that was thought of.
Such information could enable communication for severely paralyzed users or the control of a prosthesis~\cite{Pfurtscheller2008}. 

Recognizing these patterns, however, is still susceptible to errors. 
The non-stationarity and high inter-subject variance of the EEG signal requires large amounts of labeled data, which mostly is not available. 
The lack of sufficient labeled data makes training of complex classifiers with large numbers of parameters difficult. 
However, the knowledge of the structure of the EEG signal allows us to design comprehensive feature extractors, which can be combined with simpler and robust classifiers.

For extracting the most discriminative features, different approaches have been suggested. 
The well-known common spatial patterns (CSP) algorithm learns spatial filters which maximize the discriminability between two classes~\cite{LotteCSP}. 
Its performance is highly dependent on the considered operational frequency bands. 
Hence, in most applications the data is first split into several frequency bands and then spatially filtered. 
This is better known as filter bank common spatial pattern (FBCSP)~\cite{Kai_2008_winner_bci}. 

Augmented CSP features originating from a multilevel frequency decomposition have been used in Convolutional Neural Networks (CNNs) for classification~\cite{YangHuijuan2015Otuo} and achieved an average classification accuracy of 69.27\% on the 4-class data set of the BCI competition IV-2a~\cite{BCI-recording}. 
This approach is further extended in~\cite{parallel_CNN}, where multiple CNNs are fed with dynamic energy features and combined with a static energy network. 
A multilevel spectral and temporal decomposition in connection with CSPs is used as feature extractor. 
The best frequency bands and spatial filters are selected using mutual information as feature selection measure. 
With this architecture, an average classification accuracy of 70.60\% has been achieved. 

Recently, Riemannian approaches~\cite{Lotte_Riemann2017} allow the direct manipulation of spatial EEG signal covariance matrices using the dedicated Riemannian geometry.
In contrast to the Euclidean geometry, it introduces a more accurate approximation of the distance on smoothly curved spaces. 
The Riemannian approach yields a large number of features as opposed to the CSP that induces a reduction of the feature space susceptible to loss of important spatial information. 
Furthermore, Riemannian kernel features do not require labeled data for training, in contrast to CSPs, and therefore allow unsupervised feature calibration. 
Riemannian features have been classified with multi-kernel relevance vector machine achieving an accuracy of 70.30\% ~\cite{NGUYEN20181871}.

In this paper, CSP and Riemannian methods are enhanced to multiscale spectral and temporal features capturing the dynamic nature of the EEG signals.
Therefore, we introduce a new architecture which includes four stages: temporal division, spectral division, CSP or Riemannian feature generation, and classification with a support vector machine (SVM).  
This vastly increases the number of features, introduces redundancy, yet increases the classification accuracy on average by $\approx5\%$ compared to the state-of-the-art method~\cite{parallel_CNN}. 
Using almost the equal size of features, the Riemannian covariance features perform slightly better than multiscale CSP features in accuracy (74.27$\pm$15.5\% vs. 73.70$\pm$15.90\%) and execute 9$\times$ faster in training and 4$\times$ faster in testing.
Using more temporal windows increases Riemannian covariance features by 3$\times$ resulting in 75.47$\pm$12.8\% accuracy and 1.6$\times$ faster testing than CSP.
This improvement in compute time is particularly important when targeting an online, real-time implementation.   

In the following sections, we briefly introduce the CSP and Riemannian frameworks. The different feature extraction methods are tested on the BCI competition IV-2a data using a support vector machine (SVM) as classifier. 

\begin{figure*}
\vspace*{-0.36in}
\subfloat[][Multiscale CSP features using 11 temporal windows $T_1$--$T_{11}$ and 43 frequency bands $b_1$--$b_{43}$. Every CSP block uses 24 separate spatial filters and calculates the features according to \eqref{eq:csp_feat} which gives a total of 11352 features.]{ 
\includesvg[width = 0.45\textwidth]{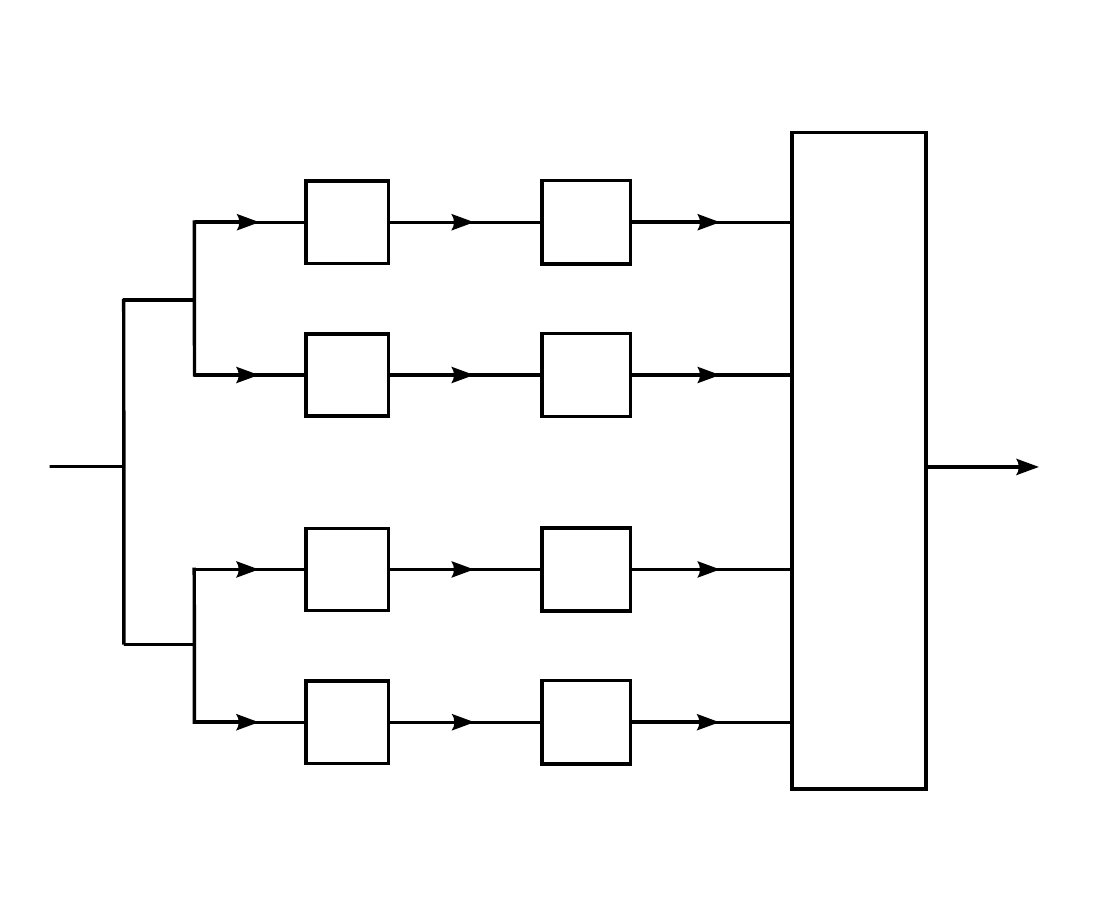}
\label{fig:tfcsp}
}
\hspace{0.05\textwidth}
\subfloat[][Riemannian covariance features using 43 frequency bands $b_1$--$b_{43}$ and one temporal window $T_1$. Every Riemannian block R includes the calculation of $\tilde{\boldsymbol{S}}$ \eqref{eq:riemann_feat} and its vectorization resulting in 253 features per block and a total of 10879 features.]{
\includesvg[width = 0.45\textwidth]{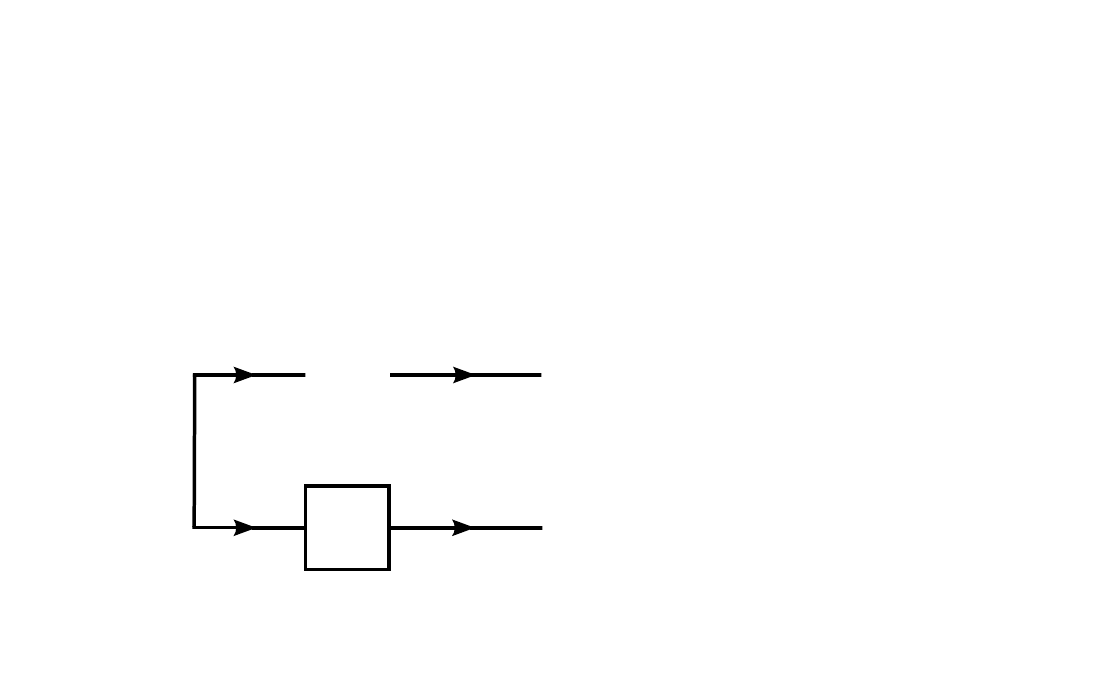}
\label{fig:tfriemann}
}
\caption{Multiscale feature generation for (a) CSP, and (b) Riemannian features. The signal is split into temporal windows as in Figure \ref{fig:temp_wind} and then band pass filtered using bands according to Figure \ref{fig:freq_bands}. All features generated either by CSP or Riemannian are stacked into one vector and fed to a SVM for classification.}
\end{figure*}

\section{Dataset Description}\label{chap:dataset}
The BCI Competition IV 2a dataset~\cite{BCI-recording} consists of EEG data from 9 different subjects. The subjects were requested to carry out four different MI tasks, namely the imagination of the movement of the left hand, right hand, both feet and tongue. Two sessions were recorded on two different days. For each subject a session consists of 72 trials per class yielding 288 trials in total. One session is used for training and the other for testing exclusively. The signal was recorded using 22 EEG electrodes according to the 10-20 system. It is bandpass filtered between 0.5\,Hz and 100\,Hz and sampled with 250\,Hz. In addition to the 22 EEG channels, three Electrooculography (EOG) channels give information about the eye movement. An expert marked the trials containing artifacts based on the EOG signal. This way 9.41\% of the trials were excluded from the dataset. The number of trials per class remains balanced.

\section{Common Spatial Pattern Features}
Common spatial pattern was first introduced for discriminating between two different kinds of populations, one with a neurological disorder and one without \cite{Koles1990}. The basic idea of CSP is to combine the spatial channels such that the average variance between two classes is maximized. We first introduce CSP for the basic binary class case. 

Given the signal $\boldsymbol{X} \in \mathbb{R}^{N_c \times N_s}$, we can estimate the covariance matrix of the signals as
\begin{align}
\boldsymbol{C} = \frac{1}{N_s-1} (\boldsymbol{X} \boldsymbol{X}^T), 
\end{align}
where $N_c$ denotes the number of channels, $N_s$ the number of time samples and $\boldsymbol{A}^T$ the transpose of a matrix $\boldsymbol{A}$. 
The signal has zero-mean component since it is bandpass filtered. 
We calculate the arithmetic average covariance matrix over all occurrences of class $j \in \{0,1\}$ by 
\begin{align}
\overline{\boldsymbol{C}}_j = \frac{1}{N_j} \sum_{k = 1}^{N_j}\boldsymbol{C_j}^{(k)},
\end{align}
where $\boldsymbol{C}^{(k)}_j$ corresponds to the covariance matrix where class $j$ occurred and $N_j$ to the total number of occurrences of class $j$. The goal of CSP is to find a spatial filter $\boldsymbol{w} \in \mathbb{R}^{N_c}$ which maximizes the Rayleigh quotient 
\begin{align}
J(\boldsymbol{w}) =  \frac{\boldsymbol{w}^T \overline{\boldsymbol{C}}_1 \boldsymbol{w}}{\boldsymbol{w}^T \overline{\boldsymbol{C}}_2 \boldsymbol{w}}.
\end{align}
This is achieved by solving the generalized eigenvalue decomposition (GEVD) problem 
\begin{align}
\overline{\boldsymbol{C}}_1 \boldsymbol{U} = \boldsymbol{\Sigma} \overline{\boldsymbol{C}}_2 \boldsymbol{U} 
\end{align}
where $\boldsymbol{\Sigma}$ is a diagonal matrix containing the eigenvalues in descending order and $\boldsymbol{U}$ contains the corresponding eigenvectors as columns. We then take pairs of eigenvectors with the corresponding largest and smallest eigenvalues. They build a set of spatial filters. Finally, the feature $f_l$ is the logarithm of the spatial filtered and normalized variances.
\begin{align}
f_l = \textrm{log}\left( \frac{\boldsymbol{w}_l^T \boldsymbol{X} \boldsymbol{X}^T \boldsymbol{w}_l}{\sum_k \boldsymbol{w}_k^T \boldsymbol{X} \boldsymbol{X}^T \boldsymbol{w}_k} \right)\label{eq:csp_feat}
\end{align}

Various approaches to extend CSP to the multiclass problem have been proposed. 
A heuristic but most successful technique is to perform two-class CSP on all possible combinations of classes~\cite{Grosse-Wentrup2008}. In the four class case, this results in at least 12 spatial filters, namely two for each pair of classes:
\begin{align} 
\lbrace (1,2), (1,3), (1,4), (2,3), (2,4), (3,4) \rbrace 
\end{align}

\subsection*{Multiscale Temporal and Spectral CSP Features}\label{sec:freq_time}
MI activities cause brain oscillations mainly within the $\mu$ (8--14\,Hz) and $\beta$ (14--30\,Hz) bands~\cite{PFURTSCHELLER1999}. 
However, which bands reveal the most discriminative information is highly subject-dependent. Furthermore, not every subject needs the same amount of time to initiate MI after the cue. 
Therefore, we divide the signal into multiscale temporal and spectral components before applying CSP. 
Figure \ref{fig:tfcsp} shows the tree-like structure of feature generation. 
We add temporal information by dividing the data into temporal windows, shown in Figure \ref{fig:temp_wind}. 
Every temporal window is split again into various spectral bands, shown in Figure \ref{fig:freq_bands}, using second order Butterworth filters. 
Both the temporal windows and the spectral bands induce redundant information which help to increase the robustness of the feature extractor. 
For every leaf we use a separate set of 24 spatial filters and apply \eqref{eq:csp_feat}.  
This results in a total number of features: 
\begin{align}
N &= N_{temp} \times N_{spec} \times N_{spat} = 11 \times 43 \times 24 = 11352
\end{align}
All the features are stacked into one vector and fed to a $\ell_2$-regularized SVM. 

\section{Riemannian Covariance Features}\label{sec:riemannian}
This section describes briefly the Riemannian geometry in the context of covariance matrices. Then, a Riemannian based kernel is introduced which can be applied on a linear SVM using the kernel trick. 
\subsubsection{Introduction to Riemannian geometry}
Roughly speaking, Riemannian geometry studies smoothly curved spaces that locally behave like Euclidian spaces \cite{Lotte_Riemann2017}. The idea is to locally approximate a smooth curved space on a tangent space. This concept is not new, since we approximate the Earth locally as a flat space. 

The space of real symmetric positive definite $N_{c}\times N_{c}$ covariance matrices $P_{N_{c}}$ forms a smooth \textit{differential manifold}, which allows the mapping between the manifolds and their corresponding local tangent space. If the number of time samples for estimating the covariance matrices is large enough, they are in $P_{N_{c}}$.

Let us fix a reference point $\boldsymbol{C}_{ref} \in P_{N_c}$. The corresponding tangent space is then denoted as $T_{C_{ref}} P_{N_c}$. As we are working with a smooth space, slightly changing the reference point $\boldsymbol{C}_{ref}$ should not affect the calculations in the tangent space too much. The logarithmic map is used to project the vectors from the submanifold $P_{N_c}$ to its tangent space $T_{C_{ref}} P_{N_{c}}$. Conversely, the exponential map projects the point on the tangent space back to the submanifold:
\begin{align} 
\boldsymbol{S} = \textrm{Log}_{C_{ref}}(\boldsymbol{C}) = \boldsymbol{C}_{ref}^{1/2}\textrm{logm}\left(\boldsymbol{C}_{ref}^{-1/2}\boldsymbol{C} \boldsymbol{C}_{ref}^{-1/2}\right) \boldsymbol{C}_{ref}^{1/2}\label{eq:riem_logmap}, \\
\boldsymbol{C} = \textrm{Exp}_{C_{ref}}(\boldsymbol{S}) = \boldsymbol{C}_{ref}^{1/2}\textrm{expm}\left(\boldsymbol{C}_{ref}^{-1/2}\boldsymbol{S} \boldsymbol{C}_{ref}^{-1/2}\right) \boldsymbol{C}_{ref}^{1/2},
\end{align}
where logm() and expm() denote the matrix logarithm and matrix exponential function, respectively. 
Next, we introduce various geometric measures in the Riemannian sense and compare them to the Euclidean counterparts. The Euclidean matrix inner product in the space of $N_{c} \times N_{c}$ real matrices is the Frobenius inner product defined as
\begin{align}
\langle \boldsymbol{C}_A,\boldsymbol{C}_B \rangle_F = \mathrm{Tr} \left(\boldsymbol{C}_A^T \boldsymbol{C}_B \right),
\end{align}
where $\mathrm{Tr}(\cdot)$ stands for the trace. Let $\boldsymbol{S}_1$ and $\boldsymbol{S}_2$ be two elements of the tangent space $T_{C_{ref}} P_{N_{c}}$. Then, we define the Frobenius inner product on the tangent space $T_{C_{ref}} P_{N_{c}}$ as
\begin{align}
\langle \boldsymbol{S}_1,\boldsymbol{S}_2 \rangle _{C_{ref}} = \mathrm{Tr} \left(\boldsymbol{C}_{ref}^{-1}\boldsymbol{S}_1\boldsymbol{C}_{ref}^{-1} \boldsymbol{S}_2 \right).  \label{eq:innprod_riem} 
\end{align}
When using two elements $\boldsymbol{C}_1$ and $\boldsymbol{C}_2$ directly out of $P_{N_{c}}$ they first need to be transformed into the tangent space using the logarithmic mapping. The inner product can be computed according to \eqref{eq:innprod_riem}. Depending on the chosen inner product we can also define different distance metrics. Let us first recall the Euclidean distance between two matrices:
\begin{align}
\delta_E(\boldsymbol{C}_A,\boldsymbol{C}_B) &= \|\boldsymbol{C}_A - \boldsymbol{C}_B \|_F 
\end{align}
Conversely, the Riemannian distance between $\boldsymbol{C}_A$ and $\boldsymbol{C}_B \in P_{N_{c}}$ is  
\begin{align}
\delta_R(\boldsymbol{C}_A,\boldsymbol{C}_B) &= \|\textrm{logm}\left(\boldsymbol{C}_A^{-1}\boldsymbol{C}_B \right) \|_F. 
\end{align}
The difference between these two distances is that the Euclidean distance determines the shortest distance along direct paths, whereas the Riemannian distance rather searches the shortest path along geodesics \cite{Riemann_mean2005}. 

Finally, the last measure we encounter often in BCI are means of covariance matrices. Lets say we have a set $\lbrace \boldsymbol{C}_i \rbrace_{i=1}^n$ including $n$ covariance matrices. The arithmetic mean which is associated with the Euclidean distance metric is 
\begin{align}
\mathfrak{U}\left(\boldsymbol{C}_1,\boldsymbol{C}_2,...,\boldsymbol{C}_n\right) = \frac{1}{n}\sum_{i=1}^n \boldsymbol{C}_i. 
\end{align}
In the Riemannian framework, we use the geometric mean (also called Fr\'{e}chet mean or Karcher mean), which tries to find a point $\boldsymbol{C}$ that minimizes the sum of all squared Riemannian distances \cite{Riemann_mean2005}
\begin{align}
\mathfrak{G}\left(\boldsymbol{C}_1,\boldsymbol{C}_2,...,\boldsymbol{C}_n\right) =  \operatornamewithlimits{argmin}_{\boldsymbol{C}} \sum_{i=1}^n \delta_R(\boldsymbol{C},\boldsymbol{C}_i)^2. 
\end{align}
This mean is not straightforward to calculate, as no closed-form solution exists. However, numerous iterative algorithms solve this problem numerically \cite{Riemann_mean2005}. 

\subsubsection{Vectorization of symmetric matrices}
To use a classifier like SVM in connection with covariance matrices, we vectorize the matrix as
\begin{align}
\overrightarrow{\boldsymbol{C}} := \textrm{vect}(\boldsymbol{C}) = [\boldsymbol{C}_{1,1};\sqrt{2} \boldsymbol{C}_{1,2}; ... \boldsymbol{C}_{N_c,N_c}] \in \mathbb{R}^{(N_c+1)N_c/2}, \label{eq:half_vec}
\end{align}
since covariance matrices are symmetric. The off-diagonal elements are scaled by $\sqrt{2}$ to preserve the norm, i.e. 
\begin{align}
\|\boldsymbol{C} \|_F = \|\textrm{vect}(\boldsymbol{C}) \|_2, \label{eq:half_vec_norm}.
\end{align} 
\subsubsection{Kernel approach}
Now we build a matrix kernel which takes the Riemannian distance into account~\cite{Barachant2013}. 
The kernel formulation of the SVM is
\begin{align}
f(\boldsymbol{C}) = \beta + \sum_{i=1}^n \alpha_i y_i k(\boldsymbol{C}_i,\boldsymbol{C};\boldsymbol{C}_{ref}). 
\end{align}
Here, $k(.,.;\boldsymbol{C}_{ref})$ is the kernel function given the reference $\boldsymbol{C}_{ref}$, $\beta$ and $\alpha_i$ trainable coefficients and $\boldsymbol{C}_i$ the support vector.
The kernel function can be written as
\begin{align}
k(\boldsymbol{C}_i,\boldsymbol{C};\boldsymbol{C}_{ref}) &= \langle \Phi (\boldsymbol{C}_i),  \Phi (\boldsymbol{C}) \rangle_{\boldsymbol{C}_{ref}} \label{eq:riem_kernel1} \\
				&= \textrm{Tr}\left(\boldsymbol{C}_{ref}^{-1}\textrm{Log}_{\boldsymbol{C}_{ref}}(\boldsymbol{C}_i)\boldsymbol{C}_{ref}^{-1}\textrm{Log}_{\boldsymbol{C}_{ref}}(\boldsymbol{C}_i)\right)\label{eq:riem_kernel2} \\
				&= \langle \tilde{\boldsymbol{S}_i},\tilde{\boldsymbol{S}} \rangle_F \label{eq:riem_kernel4} \\
				&= \overrightarrow{\tilde{\boldsymbol{S}_i}}^T\overrightarrow{\tilde{\boldsymbol{S}_i}}. \label{eq:riem_kernel5}
\end{align}
We use the Riemannian inner product $\langle .,. \rangle_{\boldsymbol{C}_{ref}}$ with respect to the reference point $\boldsymbol{C}_{ref}$ defined in \eqref{eq:innprod_riem}. Therefore, the matrices $\boldsymbol{C}_i$ and $\boldsymbol{C}$ are projected from the submanifold to the tangent space $T_{C_{ref}} P_{N_{c}}$ using the logarithmic map $\Phi (\boldsymbol{C}) = \textrm{Log}_{C_{ref}}(\boldsymbol{C})$.
We introduce a new matrix 
\begin{align}
\tilde{\boldsymbol{S_i}} = \boldsymbol{C}_{ref}^{-1/2}\textrm{Log}_{\boldsymbol{C}_{ref}}(\boldsymbol{C}_i)\boldsymbol{C}_{ref}^{-1/2}=\textrm{logm}\left(\boldsymbol{C}_{ref}^{-1/2}\boldsymbol{C}_i\boldsymbol{C}_{ref}^{-1/2} \right). \label{eq:riemann_feat}
\end{align}
Together with the definition of the Frobenius norm on the Euclidean space, we obtain \eqref{eq:riem_kernel4}. Due to the norm conversation \eqref{eq:half_vec_norm} of the vectorization, we get \eqref{eq:riem_kernel5}~\cite{Barachant2013}. 

This equivalence allows a kernel-free linear SVM implementation. The reference matrix $\boldsymbol{C}_{ref}$ can be selected in various ways. 
For example, one could use a mean covariance matrix (either the geometric $\mathfrak{G}$ or arithmetic $\mathfrak{U}$ mean) from the training data or simply a ${N_{c}} \times {N_{c}}$ identity matrix $\mathfrak{I}$. 
Using the identity matrix as reference is equivalent to calculating the matrix logarithm and applying the vectorization \eqref{eq:half_vec}. 
Finally, we apply the same multiscale division as in the previous CSP section. However, since the number of features per leaf is 253, we use only the largest temporal window. 
The structure is shown in Figure~\ref{fig:tfriemann}.
A separate reference point $\boldsymbol{C}_{ref}$ is used for every frequency band and is averaged over all temporal windows. 

\begin{table*}
\vspace{0.05in}
\centering
\caption{Classification accuracy (\%), training and testing time for the common spatial patterns (CSPs) and Riemannian features tested with different SVM kernels.}
\label{tab:results}
\begin{tabular}{lrrrrrrrr}
\toprule
               	& \multicolumn{4}{c}{CSP} 				& 	\multicolumn{4}{c}{Riemannian} \\
\cmidrule(r){1-1}\cmidrule(r){2-5}\cmidrule(r){6-9}
No. features 	& 11352		 & 11352	& 11352			&  20856   	& 10879		& 10879		& 10879	& 32637 	\\ 
SVM kernel		& linear & rbf			& poly 			& linear  	&linear 	&linear 	&linear &linear   \\ 
Riemannian kernel &N/A  & N/A 			& N/A 		& N/A  	&$\mathfrak{G}$&$\mathfrak{U}$&$\mathfrak{I}$ &$\mathfrak{G}$   \\ 
Spectral bands 	&$b_1$--$b_{43}$ & $b_1$--$b_{43}$	& $b_1$--$b_{43}$ 		& $b_1$--$b_{80}$  	&$b_1$--$b_{43}$	&$b_1$--$b_{43}$	&$b_1$--$b_{43}$ &$b_1$--$b_{43}$   \\ 
Temporal windows	&$T_1$--$T_{11}$  & $T_1$--$T_{11}$ 			& $T_1$--$T_{11}$		& $T_1$--$T_{11}$  	&$T_1$	&$T_1$	&$T_1$ &$T_1$,$T_2$,$T_5$   \\ 
\cmidrule(r){1-1}\cmidrule(r){2-5}\cmidrule(r){6-9}
Subject 1       & 86.83  &  85.41		& 83.27        	& 84.70   	& 91.81	    & 90.75		& 84.70	& 90.04\\
Subject 2       & 57.24  & 57.24		& 49.47        	& 57.60   	& 51.59     & 47.70		& 48.76	& 55.48\\
Subject 3       & 86.45  & 80.95		&77.66        	& 84.98   	& 83.52     & 85.35		& 84.25	& 81.32\\
Subject 4       & 61.40  & 62.28		&56.58      	& 58.33   	& 73.25		& 63.16		& 58.33	& 71.93\\
Subject 5       & 61.23  & 67.03		&60.14        	& 63.77   	& 63.41		& 67.39		& 62.32	& 69.57\\
Subject 6       & 50.70  & 50.23		&44.65      	& 48.84   	& 58.60		& 58.60		& 56.74	& 56.74\\
Subject 7       & 92.42  & 86.28		&70.40	        & 88.45  	& 86.64		& 89.89		& 81.59	& 85.56\\
Subject 8       & 87.82  & 87.45		&86.72	        & 87.08   	& 81.55		& 85.24		& 79.34	& 83.76\\
Subject 9       & 79.17  & 85.98		&79.92        	& 84.85   	& 82.58		& 80.30		& 80.68	& 84.85\\ 
 \cmidrule(r){1-1}\cmidrule(r){2-5}\cmidrule(r){6-9}
Avg. accuracy	&\textbf{73.70}$\pm$15.9  & 73.65$\pm$14.5 &67.65$\pm$15.4	&73.18$\pm$15.7 &\textbf{74.77}$\pm$13.9 	&74.27$\pm$15.5		& 70.75$\pm$14.0 &  \textbf{75.47}$\pm$12.8\\ 
Avg. training time [s] &49.46&42.19			&  43.65			& 	97.77		& 17.08		& \textbf{5.45}		& 13.79	& 50.93 \\
Avg. testing time [s]  &23.19&24.24 			& 25.40			&	47.35		& \textbf{5.53}		& \textbf{5.33}		& 5.56	& 13.88 \\ \bottomrule

\end{tabular}
\end{table*}

\section{Experimental Results}
The aforementioned approaches are evaluated on the dataset 2a of the BCI competition IV. The classification accuracy is measured as
\begin{align}
\textrm{classification accuracy} = \left( \frac{N_{\textrm{correct}}}{N_{\textrm{total}}} \right) \times 100 \%, 
\end{align} 
where $N_{\textrm{correct}}$ is the number of correct classified trials and $N_{\textrm{total}}$ the total number of trials in the test set per subject.
To evaluate the computational cost the average training and testing time per subject over all trials is measured. 
The training time includes the preprocessing and the training of the classifier, whereas the testing time covers the calculation of the features as well as the classification itself. 
The experiments were conducted on an Intel Core i7-4790 3.6\,GHz processor with 16\,GB RAM. The code is available\footnote{\url{https://github.com/MultiScale-BCI/IV-2a}}. 

The features are classified with a $\ell_2$-regularized SVM. 
The regularization hyperparameter $C$ is determined in 5-fold cross-validation using grid search within the range $[10^{-2},10^3]$. Moreover the temporal windows and frequency bands were selected in cross-validation as well. Table \ref{tab:results} shows the results on the test set. 

The proposed Riemannian covariance features achieve higher accuracies than CSPs with a similar number of features. 
Moreover, the testing time of the Riemannian features is 4$\times$ lower than those of the CSP. 
The Riemannian feature extractor exploits 253 features per frequency band and temporal window instead of 24 in the CSP case. A single temporal window is sufficient for the Riemannian features to achieve the same accuracy as the CSPs.

We evaluate different SVM kernels for the multiscale 11352 CSP features. 
The linear kernel achieves 73.70\% classification accuracy. 
The nonlinear radial basis kernel function classifies on par with the linear kernel while the polynomial kernel was outperformed. 
%
%


For comparing the geometric mean $\mathfrak{G}$, the arithmetic mean $\mathfrak{U}$ and the identity matrix $\mathfrak{I}$ as reference point in the Riemannian kernel, we use the calculated features of one temporal window ($T_1$) and 43 spectral bands ($b_1$--$b_{43}$) yielding 10879 features. 
All means are calculated on the training set in order to keep the feature extractor causal. A difference in the training time between the reference point calculation methods is observed: the geometric mean requires iterative numerical operations which results in 3$\times$ higher training time than the arithmetic mean. In terms of accuracy, however, we gain 0.50\% accuracy by using the geometric mean. 

For the highly subject dependent nature of this classification task, we use large multiscale features. However, increasing the number of features to 20856 by adding 1\,Hz frequency bands was not found to be beneficial for CSP.
Adding more temporal windows ($T_2$ and $T_5$) to the Riemannian improves classification accuracy (75.47$\pm$12.8\%) at the cost of 3$\times$ higher computation time. 
Using even more temporal windows reduces the accuracy that can be due to the overfitting situation for more features with the limited training samples.

\section{Conclusion}
We propose an enhanced feature extraction method based on multiscale temporal windows and overlapping spectral bands for multiclass classification of EEG signals.
Our method significantly increases the number of features for the CSP and Riemannian covariance methods. 
In combination with a simple linear SVM classifier our method outperforms the state-of-the-art on average by $\approx5\%$. 
This confirms the importance of large multiscale temporal and spectral features for the MI-BCIs. 
Besides, the Riemannian covariance method is unsupervised, and achieves 4$\times$ faster execution time during testing (compared to CSP with almost equal number of features). 
These are particularly important for a real-time embedded implementation.

\section*{Acknowledgments}
Support was received from the ETH Zurich Postdoctoral Fellowship program and the Marie Curie Actions for People COFUND Program.

\bibliographystyle{IEEEtran} 
\bibliography{bibliography}

\end{document}